# Ion Cyclotron Resonance Heating Systems Upgrade toward High Power and CW operations in WEST

Julien Hillairet[1,a], Patrick Mollard[1], Yanping Zhao[2], Jean-Michel Bernard[1], Yuntao Song[2], Arnaud Argouarch[1], Gilles Berger-By[1], Nicolas Charabot[1], Gen Chen[2], Zhaoxi Chen[2], Laurent Colas[1], Jean-Marc Delaplanche[1], Pierre Dumortier[4], Frédéric Durodié[4], Annika Ekedahl[1], Nicolas Fedorczak[1], Fabien Ferlay[1], Marc Goniche[1], Jean-Claude Hatchressian[1], Walid Helou[1], Jonathan Jacquot[1], Emmanuel Joffrin[1], Xavier Litaudon[1], Gilles Lombard[1], Riccardo Maggiora[3], Roland Magne[1], Daniele Milanesio[3], Jean-Claude Patterlini[1], Marc Prou[1], Jean-Marc Verger[1], Robert Volpe[1], Karl Vulliez[6], Yongsheng Wang[2], Konstantin Winkler[5], Qingxi Yang[2], Shuai Yuan[2]

1. CEA, IRFM, F-13108 Saint Paul-lez-Durance, France.
2. Institute of Plasma Physics (ASIPP), Chinese Academy of Sciences, Hefei 230031, China
3. Department of Electronics, Politecnico di Torino, Torino, Italy
4. Laboratoire de physique des plasmas de l'ERM, Laboratorium voor plasmafysica van de KMS – (LPP-ERM/KMS) - Ecole royale militaire-Koninklijke militaire school - BE-1000 Brussels-Belgium
5. Max-Planck Institut für Plasmaphysik, Boltzmannstraße 2, 85748 Garching, Germany
6. Laboratoire d'étanchéité, DEN/DTEC/SDTC, CEA, 2 rue James Watt 26700 Pierrelatte, France

a) Corresponding author: julien.hillairet@cea.fr

**Abstract.** The design of the WEST (Tungsten-W Environment in Steady-state Tokamak) Ion cyclotron resonance heating antennas is based on a previously tested conjugate-T Resonant Double Loops prototype equipped with internal vacuum matching capacitors. The design and construction of three new WEST ICRH antennas are being carried out in close collaboration with ASIPP, within the framework of the Associated Laboratory in the fusion field between IRFM and ASIPP. The coupling performance to the plasma and the load-tolerance have been improved, while adding Continuous Wave operation capability by introducing water cooling in the entire antenna. On the generator side, the operation class of the high power tetrodes is changed from AB to B in order to allow high power operation (up to 3 MW per antenna) under higher VSWR (up to 2:1). Reliability of the generators is also improved by increasing the cavity breakdown voltage. The control and data acquisition system is also upgraded in order to resolve and react on fast events, such as ELMs. A new optical arc detection system comes in reinforcement of the $V_r/V_f$ and SHAD systems.

## INTRODUCTION

The mission of WEST (Tungsten-W Environment in Steady-state Tokamak) is to mitigate the risks associated with the procurement of the ITER divertor and gaining time for its operation [1]. It consists of testing an ITER like actively cooled tungsten divertor with realistic ITER load case conditions, i.e. with heat flux of 10 MW/m² for up to thousand seconds. To fulfil these aims, the Tore Supra tokamak is transformed into an X-point divertor device with the capability to run long pulses in H-mode. Two RF heating and current drive systems are available in WEST: Ion cyclotron resonance heating (ICRH) and Lower Hybrid Current Drive [2]. The main scenario allowing to reach ITER relevant heat fluxes relies on three ICRH antennas, designed to operate for the first plasmas in Hydrogen minority heating scheme up to 3 MW each for 30 s and then at 1 MW each for 1000 s. Edge Localized Modes (ELMs) induced by H-mode operations are expected in WEST. ELMs induce strong and fast quasi-periodic plasma edge density variations, resulting in load variations for the ICRH antenna. ELM resilient system has to be employed in order to reduce the impact of these variations on the generators. Moreover, long pulses requires Continuous Wave

(CW) RF systems. The WEST ICRH system is both ELMs resilient and CW, two challenging issues that no other ICRH system before ITER has faced simultaneously so far, moreover in a fully metallic environment.

This paper gives an overview of ICRH system upgrades. The first section details the generator plant upgrades. The second section reviews the antennas RF design and the trades-off made in order to increase the antenna coupling efficiency, while ensuring its impedance matching capabilities and CW operations in metallic environment. The third section lists the auxiliary system and diagnostic upgrades.

## GENERATOR PLANT UPGRADES

The ICRH plant is made of three modules (one per antenna) of two generators each (one per half-antenna). The six generators are identical and consist of a synthesizer, a modulator, a solid-state wideband amplifier and a three-stage Thales Electron Devices tetrode amplifier. The plant can operate in the 40-78 MHz frequency range. The nominal operating frequencies are $53 \pm 2$ MHz and $55.5 \pm 2$ MHz in order to allow some flexibility in the localization of the resonance layer. Considering a 30 s pulse, the high power target per generator is 1 MW with a VSWR below 1.5:1. The high power limitation comes from the maximum anode dissipation or the maximum allowed anode voltage, which are specifics to each tube. The generators can operate during 1000 s at 500 kW, independently of the VSWR. The CW power limitation comes from the RF losses in the final stage of the tetrode. It has to be noted that the 9" 30 Ohm transmission lines are not cooled and thus are limited to 800 kW CW in order to keep the inner conductor temperature below 100°C. Limitations also exist with the antenna vacuum feedthroughs in CW (500 kW) or high power (1.75 MW/50 s).

The anode decoupling capacitors of all the generators final stage have been upgraded internally following TED specifications in order to increase the cavity breakdown voltage limit and thus the reliability during high power fast VSWR variations. The modification consists in doubling the distance between the anode and the screen grid decoupling capacitors.

Even with a load resilient system (3 dB coupler or load-tolerant electrical scheme), generators are affected by ELM fronts which are typically faster (<100 µs) than generator power regulation response time, resulting in unstable power during ELMs. In order to insure additional margins with respect to ELMy plasma load variations, the operation class of the generators will be changed from AB to B, in order to allow more power from the generator at a given VSWR-value, at the expense of a slightly reduced gain. In addition, the electronic regulation has been upgraded with FPGA system to improve the response time and reduce the number of disjunctions.

## LOAD-RESILIENT CW ANTENNAS

Three new identical ELM-resilient and CW power ICRH antennas have been designed for WEST. The ELM resilience property is obtained through an internal conjugate-T electrical scheme [3] with series capacitors. The antenna design is based on a previously tested prototype at 48 MHz [4,5]. In the frame of the Associated Laboratory in the fusion field between IRFM and ASIPP, the design has been upgraded in order to: i) improve the power capabilities ii) increase the nominal frequency range to 55 MHz [6] iii) unlike the prototype, allow CW operation with actively cooled components [7]. This latter aspect led to iterate between RF and mechanical design, with generally opposite constraints and necessary compromises. In addition, it shall be noted that a compromise has been often required between the increase of the straps coupling resistance on the one hand and the decrease of their reactance on the other hand [6].

A specific attention has been paid to the antenna front face. The coupling characteristics have been calculated in COMSOL® using approximate 2D and 3D parametrized models of the prototype antenna front face, facing a cold-plasma surface or volume with appropriate boundaries conditions [8]. The optimization process has been performed with reference reflectometry electron density profiles measured in 2007 during the Tore Supra pulse #40574. In this pulse, an L-mode plasma has been moved horizontally over 5 cm in front of the load-resilient antenna prototype to assess its coupling properties at 48 MHz. The calculations have been performed for 8 profiles in order to make relative comparisons of the coupling resistance. 2D model runs in few minutes while 3D models typically require ten to twenty hours depending on their complexity and the distance to the cut-off density on a desktop computer. These parametrized models allowed us to assess the relevant geometrical parameters (strap thickness, width, and distances to the plasma cut-off and antenna box) affecting the coupling resistance.

In order to optimize the coupling resistance in dipole configuration, the thickness of the strap has been kept as low as possible (between 13 and 15 mm) while ensuring sufficient stiffness, since they are water cooled through

70°C/30 bars inner channels. As an example, at 57 MHz, increasing the strap thickness by 15 mm decreased the coupling resistance by 25% for the same plasma loading. The coupling resistance of the new antennas has been increased compared to the prototypes one in particular by reducing the distance between the strap and the Faraday screen bars from 52 mm to 34 mm and by supressing a vertical septum which was masking a portion of the straps in the prototype.

Each antenna is equipped with four internal COMET® tuneable vacuum capacitors, with capacitances ranging from 15 pF to 150 pF and specifically upgraded for CW operation. In order to operate the antenna at a higher frequency, 55 MHz compared with the 48 MHz of the prototype, with the same capacitance range, the straps electrical length has been reduced in order to decrease their reactance. However, bringing closer the strap and its feeder leads to reduce the coupling resistance (-23% decrease of the coupling resistance for 50 mm distance reduction). Increasing the toroidal width of the straps led to reduce the strap reactance but also to decrease the coupling resistance (-50% decrease of the coupling resistance for +40 mm width increase). It has been kept to 130 mm, the same value as for the prototype. As the design of the antenna progressed, the coupling of few milestone CAD 3D models has been calculated with the TOPICA code [9] with the same plasma scenarios to assess the impact of mechanical design evolutions on the coupling performances.

The T-junction, the capacitor and the strap housings are also actively cooled, similarly to the ITER design [10]. The length of the T-junction section (or "bridge") has been increased by 50 mm in comparison to the prototype, in order to increase the available space for the water pipes and the diagnostic cables. A two-stage quarter-wavelength and water cooled impedance transformer is connected from the T-junction to the vacuum feedthrough. The impedance transformer dimensions have been optimized for a 3 Ω T-junction impedance and a 30 Ω feeder characteristic impedance at 50 MHz. The final diameters have been afterwards adjusted to match pipe diameter standards with a slight compromise with the RF performance [6].

RF circuit optimizations have been first conducted on a single conjugate-T with uncoupled straps loaded with impedances. Thereafter, full-wave models of the T-junction, impedance transformer and vacuum feedthrough have been taken into account in ANSYS Designer®. This allows pushing realistic RF excitations into the full-wave ANSYS HFSS® model in order to deduce the heat flux topologies over the devices, which then can be easily imported in ANSYS Mechanical® for thermal modelling. Capacitors have been modelled using equivalent circuits. Finally, an electric circuit of the full antenna using TOPICA impedance matrices has been simulated. This model has been used to deduce the realistic capacitor sets, current and voltage values to be expected for various plasma scenarios. A full-wave modelling of the capacitor has also been performed and integrated into a RF circuit solver to confirm the approach [11].

Expected electron density profiles for WEST H-mode plasmas have also been used to load the antenna model. A set of profiles have been calculated for 3 different line average densities (FIGURE1 (a)). Since the plasma volume in WEST is smaller than in Tore Supra, the antennas are located closer to the centre of the machine to keep the same gap from the antenna to the plasma. Three radial locations of the antenna have been used (R=3, 2.975 and 2.95 m) to model the coupling in TOPICA with these profiles, with the toroidal curvature of the plasma adjusted accordingly to the WEST magnetic ripple. As no data were available for H-mode temperature profile, L-mode profile has been used. It has to be noted than the private plasma located between the antenna limiters has not been included in the TOPICA models.

Using the coupling resistance definition from ref. [6] and the maximum capacitors currents and voltages (850 A and 54 kV peak respectively) the maximum coupled power has been calculated. In order to balance the top and bottom strap current magnitudes, one can allow the launchers input impedances to have an additional imaginary part. It might increase the VSWR but still allowing higher coupled power: an unbalanced current distribution can lead to currents/voltages exceeding the maximum limits at one strap while they are by far respecting the limits for the other one. FIGURE1 (b) illustrates the maximum coupled power achievable for the three antennas with respect to capacitors current and voltage limitations, for two matching strategies (VSWR=1:1 or VSWR<1.7:1). In this curve, the antennas have been matched for each coupling resistance point.

## AUXILIARY SYSTEMS AND DIAGNOSTICS

The antenna control and data acquisition system (CODAC) sampling period is reduced from 1 ms to few µs in order to resolve and react on fast events, such as ELMs. An FPGA-based system acquires the antenna incident and reflected powers, voltage magnitudes and phases and is in charge of safety power feedback controls. Slower acquisition systems (~1 ms) are devoted to the control of antenna radial position, capacitors control and matching algorithms, generator plant settings and vacuum interlocks. The capacitors internal vacuum is also monitored on a regular daily basis (without magnetic field).

A new optical arc detection system comes in reinforcement to the voltage ratios ($V_r/V_f$) and Sub-Harmonic Arc Detection (SHAD) systems. This system relies on 6 optical fibres per antenna associated to fast photodetectors (<10 μs), looking into the low impedance regions around the T-junction where the $V_r/V_f$ monitor is known to be not reliable [4,12,13].

Six edge reflectometry WR-28 waveguides have been integrated into each antenna, at its top, bottom and equatorial plane. Gas puffing pipes have also been added to each antenna sides, located behind the lateral limiters.

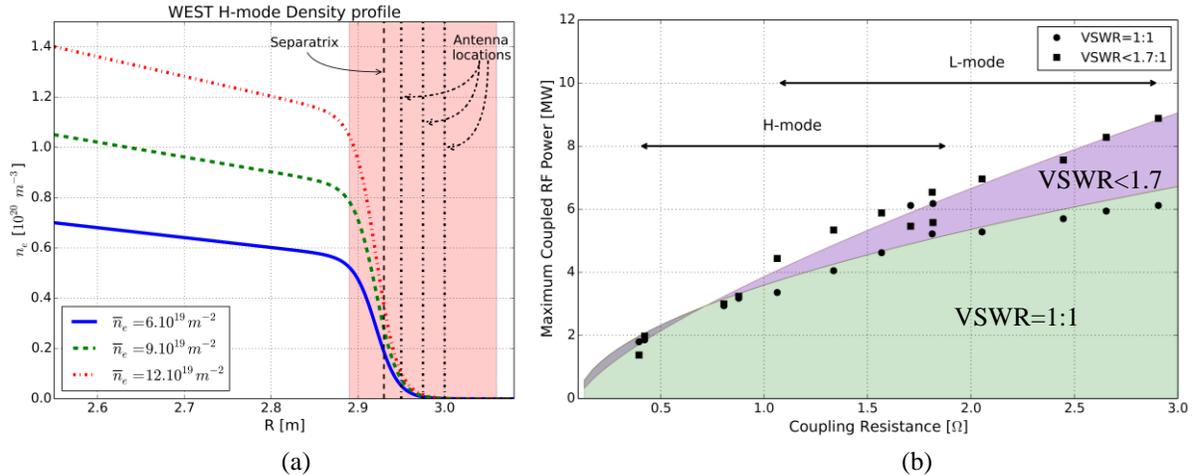

**FIGURE 1.** (a) : WEST H-mode density profiles (simulation) for 3 line average densities. The radial location of the separatrix (R=2.93 m) and of 3 antenna radial locations are indicated. The toroidal magnetic field for WEST is 3.6 T at R=2.529 m. Plasma composition 95%D-5%H. The red band corresponds to the antennas radial position range.
(b): Maximum coupled power versus coupling resistance for two matching strategies calculated from the circuit model. Curves correspond to points best fit.

## CONCLUSION

The WEST ICRH system is upgraded in order to allow both steady-state and ELM-resilience operation simultaneously. The goal is to allow the coupling of up to 9 MW total for 30 s and 3 MW total for 1000 s. The generator plant is modified in order to tolerate operation at higher VSWR. The final stage anode decoupling capacitors have been upgraded to increase the cavity breakdown voltage limit and thus the reliability during high power fast VSWR variations. Three new CW and load-tolerant antennas have been designed and are being manufactured in collaboration with ASIPP. A new optical arc detection system has been added to the antennas in order to complement the $V_r/V_f$ and SHAD systems. A new CODAC system with a fast sampling rate allows detecting fast events such as ELMs and acting to tune or protect the antennas and generator plant.

## REFERENCES

1. J. Bucalossi et al., Fusion Engineering and Design 89 (2014), 907–912.
2. R. Magne et al. AIP Conference Proceedings, 1580, 211-214 (2014).
3. G. Bosia, Fusion Sci. Technol., 43 (2003) 153–160.
4. K. Vulliez et al., Nucl. Fusion 48 (2008) 065007.
5. A. Argouarch et al., Fusion Engineering and Design 84 (2009) 275–278.
6. W. Helou et al., Fusion Engineering and Design, In Press.
7. K. Vulliez et al., Fusion Engineering and Design, In Press.
8. J. Jacquot et al, Plasma Phys. Control. Fusion 55 (2013) 115004.
9. D. Milanesio et al., Nucl. Fusion 49 (2009) 115019.
10. K. Winkler et al., AIP Conference Proceedings 1580, 366 (2014).
11. W. Helou et al., this conference.
12. P. Dumortier et al., Fusion Engineering and Design, 86 (2011) 831-834.
13. G. Berger-By et al., Fusion Engineering and Design, 82 (2007) 716-722.